\documentclass[twocolumn,
showpacs,preprintnumbers]{revtex4}
\usepackage{graphicx}

\begin{document}
\draft
\preprint{\vtop{{\hbox{YITP-05-01}
}}}
\thispagestyle{empty}
\title{Charmed scalar resonances \\
--- Conventional and four-quark mesons
}
\author{K. Terasaki}\email{terasaki@yukawa.kyoto-u.ac.jp}
\affiliation{ Yukawa Institute for Theoretical Physics, 
Kyoto University, Kyoto 606-8502, Japan
}
\author{Bruce H J McKellar}\email{mckellar@physics.unimelb.edu.au}
\affiliation{School of Physics, 
The University of Melbourne, Victoria 3010, 
Australia}
\date{January 19, 2005}
\thispagestyle{empty}
\begin{abstract}
We propose that there coexist two scalar mesons of different
structures (the conventional $^3P_0\,\,\{c\bar n\}$ meson 
and a scalar four-quark $[cn][\bar u\bar d]$ meson) in the 
recently observed broad bumps just below the large peak of the tensor 
meson in the $D\pi$ mass distribution.   We base this proposal on the interpretation of the $D_{s0}^+(2317)$ as a $[cn][\bar s\bar n]$ four quark meson.
The strange counterparts of these scalar mesons are also studied. 
\end{abstract}

\vskip 0.5cm
\maketitle
We propose that the broad bumps with $\sim 240 - 280$ MeV widths in the $D\pi$ invariant 
mass distributions,  seen in $B$ decays 
just below the large peak of the well-known tensor meson $D^*_2(2460)$~\cite{BELLE-D_0,FOCUS} should be re-interpreted as four quark mesons and conventional mesons.  We do so because of the evidence that the charm-strange scalar meson $D_{s0}^+(2317)$ should be regarded as a four quark state, and we expect its non-strange partners to be in this mass region below 2460 MeV.

The charm-strange scalar meson $D_{s0}^+(2317)$ recently 
observed at the B-factories~\cite{BABAR,CLEO,BELLE-D_s}, was predicted~\cite{DGG,chiral}, and has been 
variously interpreted as
the isosinglet state 
(the conventional scalar $\{c\bar s\}$~\cite{DGG} which is 
the chiral partner of $D_s^+$~\cite{chiral}),  
a scalar four-quark state~\cite{CH}, 
a $DK$ molecule~\cite{BCL} or atom~\cite{atom}, or bound state~\cite{BR}, in chiral quark models~\cite{Lutz}, as a  a diquark-antidiquark~\cite{MPPR}, 
as a mixed state of $\{c\bar s\}$ and a four-quark meson~\cite{BPP},  and as an $I=1$ four quark meson~\cite{Terasaki-D_s}.

The experimental result given by the CLEO 
collaboration~\cite{CLEO}, 
\begin{equation}
\frac{\Gamma(D_{s0}^+\rightarrow D_s^{*+}\gamma)
}{\Gamma(D_{s0}^+\rightarrow D_s^{*+}\pi^0)
}
 < 0.059, 
                                                   \label{eq:CLEO}
\end{equation}
is a severe constraint on the interpretation of the resonance, and it favours the assignment of the $D_{s0}^+(2317)$ to the 
$(I,I_3)=(1,0)$  four-quark meson, 
$\hat F_I^+\sim [cn][\bar s\bar n]_{I=1},\,(n=u,\,d)$~\cite{HT}.  
(We use the classification of the four-quark mesons of
Ref.~\cite{Terasaki-D_s}.)   This interpretation implies that, in addition to the 
conventional $\{c\bar q\}$ with $q=u,\,d,\,s$, scalar four-quark 
mesons indeed exist.   If a charm-strange meson of this type exists, we would expect to find its charm-non-strange partners.  Where are they?

In the $B \to D \pi\pi$ decays,  the $D\pi$ invariant 
mass distributions shows broad bumps with $\sim 240 - 280$ MeV widths 
just below the large peak of the $D^*_2(2460)$ tensor meson~\cite{BELLE-D_0,FOCUS}.  We note that  the results from two experiments are a little 
different from each other, and, in addition, it has been claimed in these papers
that these bumps are consistent with the conventional scalar 
$D_0^*\sim\{c\bar n\},\,\,(n=u,\,d)$ mesons.
Given the evidence that the  $D_{s0}^+(2317)$ is the $\hat F_I^+\sim [cn][\bar s\bar n]_{I=1},\,(n=u,\,d)$ four quark state, we regard this claim as unrealistic.  There would be no room for the 
non-strange counterparts, $\hat D\sim [cn][\bar u\bar d]$, 
of the four-quark meson, if 
the above bumps were saturated only by $D_0^*$ states.
In this short note,  we study decays of the $D_0^*$ 
mesons and demonstrate that the conventional $D_0^*$ and the 
four-quark $\hat D$ can co-exist in the broad bump regions.  
We also study  decays of the conventional scalar 
$D_{s0}^{*+} \sim \{c\bar s\}$ 
into its  dominant  $DK$ 
decay modes. Our results will be useful in the (re)analysis of the 
$D\pi$  and $DK$ invariant mass distributions we hope will occur in near future. 

The conventional scalar $D_0^*$'s have been expected to be in the 
region of $m_{D_0^*}\sim 2300 - 2400$ MeV from various approaches: 
for example, potential models~\cite{GK,potential}, and lattice 
QCD~\cite{quench,Bali,UKQCD}. However, the results from QCD sum 
rules are not yet stable --- in one 
case~\cite{QCDSR-HT}, the result is similar to those in 
potential models and lattice QCD, while the result is much 
lower in the other case~\cite{Narison}. The non-strange iso-doublet 
counterparts, $\hat D$'s, of the four-quark $\hat F_I^+$ meson 
have been predicted~\cite{Terasaki-D_s} to be around 
$m_{\hat D}\simeq 2.22$ GeV (near the lower tail of the broad 
bumps in the $D\pi$ mass distributions), using  simple quark 
counting with the mass difference,  
$\Delta_s =  m_s - m_n \simeq 100$ MeV.  
Their widths are expected to be about 50 \% broader than 
that of the $D_{s0}^+(2317)$ but they are still  
narrow~\cite{Terasaki-D_s,Terasaki-ws,Terasaki-D_0}. 
Therefore, we expect that two different scalar iso-doublets, 
the conventional $D_0^*$ and the four-quark $\hat D$ mesons, can
co-exist in the region of the broad bump of the $D\pi$ invariant 
mass distributions~\cite{Terasaki-D_0,Terasaki-mquark}. 

To make these arguments more precise, we first study decays of the conventional scalar $D_0^*$ mesons. 
Their widths are expected to be approximately saturated by the decays to 
 $ D\pi$ states.
Since the $K_0^*(1430)$'s have been considered as $^3P_0$ 
$\{n\bar s\}$ states~\cite{CT}, the $D_0^{*}\rightarrow D\pi$ decays 
can be compared with the $K_0^*(1430)\rightarrow K\pi$ decays. 
In general the two body decay, 
$A({\bf p}) \rightarrow B({\bf p'}) + \pi({\bf q})$, 
has the rate 
\begin{eqnarray}
\Gamma(A \rightarrow B + \pi)&&
=\Biggl({1\over 2J_A + 1}\Biggr)  
\Biggl({q_c\over 8\pi m_A^2}\Biggr)  \nonumber\\
&&\times
\sum_{spins}|M(A \rightarrow B + \pi)|^2 , 
\label{eq:rate}
\end{eqnarray}
where $J_A$, $q_c$ and $M(A \rightarrow B + \pi)$ denote the spin 
of the parent $A$, the center-of-mass momentum of the final $B$ and 
$\pi$ mesons, and the decay amplitude, respectively. To calculate 
the amplitude, we use the PCAC (partially conserved axial-vector 
current) hypothesis and a hard pion approximation in the infinite 
momentum frame, i.e., ${\bf p}\rightarrow\infty$~\cite{suppl}. 
In this approximation, the amplitude is evaluated at the slightly
unphysical point, i.e., $m_\pi^2 \rightarrow 0$. By assuming that 
the $q^2$ dependence of the amplitude is mild,  as was customary in 
current algebra~\cite{MP}, $M$ is given by
\begin{equation}
M(A \rightarrow B + \pi) 
\simeq
f_\pi^{-1} \left(m_A^2 - m_B^2\right)
\langle{B|A_{\bar \pi}|A}\rangle ,                         
\label{eq:amp}
\end{equation}
where $A_\pi$ is the axial charge, i.e., the counterpart of the isospin, 
$V_\pi$. 

The {\it asymptotic matrix element} of $A_\pi$ (or the matrix element of 
$A_\pi$ taken between single hadron states with infinite momentum), 
$\langle{B|A_\pi|A}\rangle$, gives the dimensionless $AB\pi$ 
coupling strength. We parameterize the asymptotic matrix elements 
of $A_\pi$ and $A_K$ using asymptotic flavor $SU_f(4)$ symmetry, 
which is, roughly speaking, $SU_f(4)$ symmetry of asymptotic matrix 
elements. (Asymptotic flavor symmetry and its fruitful results were reviewed 
in Ref.~\cite{suppl}.) 
We expect this asymptotic flavor symmetry to be broken, and a
 measure of the (asymptotic) flavor symmetry breaking is given 
by the form factor, $f_+(0)$, of the related vector current. 
The estimated values of $f_+(0)$'s are 
\begin{eqnarray}
f_+^{(\pi K)}(0)&&= 0.961 \pm 0.008, \label{eq:LR}\\
f_+^{(\bar K D)}(0)&&= 0.74 \pm 0.03,  \label{eq:PDG96}\\
\left[f_+^{(\pi D)}(0)\right]\left/
\left[f_+^{(\bar K D)}(0)\right]
\right.
&&= 1.00 \pm 0.11 \pm 0.02, \label{eq:E687} \\
&&= 0.99\pm 0.08,  \label{eq:CLEO97}
\end{eqnarray}
where the values in Eqs.~(\ref{eq:LR}) -- (\ref{eq:CLEO97}) have 
been taken from Refs.~\cite{LR} --  ~\cite{CLEO97}, respectively. 
They imply that the asymptotic flavor $SU_f(3)$ symmetry works 
well while the $SU_f(4)$ is broken to the extent of 20 -- 30 $\%$. 
This estimate is confirmed by the observation that  the asymptotic symmetry has predicted the 
rates~\cite{suppl,HOS}, 
$\Gamma(D^{*+}\rightarrow D^0\pi^+) \simeq 96$ keV and 
$\Gamma(D^{*+}\rightarrow D^+\pi^0) \simeq 42$ keV, 
which are larger by about $40\,\,\%$ than the observed values, 
 $\Gamma(D^{*+}\rightarrow D^0\pi^+)_{\rm exp} = 65 \pm 18$ keV 
and 
$\Gamma(D^{*+}\rightarrow D^+\pi^0)_{\rm exp} = 30 \pm 8$ keV,  
obtained from the measured decay width~\cite{CLEO-D^*},
$\Gamma_{D^{*\pm}}=96 \pm 4 \pm 22$ keV, 
and the branching fractions compiled in Ref.~\cite{PDG04}. 
The above suggests that asymptotic $SU_f(4)$ symmetry 
overestimates the size of the asymptotic matrix elements of the axial 
charge $A_\pi$ between charmed meson states by about $20\,\%$ 
compared with the measured rate, as expected from the above values 
of the form factors, $f_+(0)$'s.  
However, for simplicity, we will use 
asymptotic $SU_f(4)$ symmetry relations among asymptotic matrix
elements of $A_\pi$ and $A_K$ 
in our estimates of
decay rates. When we take account for the symmetry breaking, 
we will note it.  

We are now ready to study the $K_0^{*}\rightarrow K\pi$ decays and 
estimate the size of the asymptotic matrix element of $A_\pi$ taken 
between $\langle{K}|$ and $|K_0^{*}\rangle$. 
Substituting the measured values~\cite{PDG04}, 
$\Gamma(K_0^{*}\rightarrow all)=294\pm 23$ MeV with 
${\rm Br}(K_0^{*}\rightarrow K\pi)=93 \pm 10\,\,\%$,  
into Eq.(\ref{eq:rate}) and using Eq.(\ref{eq:amp}), we obtain 
\begin{equation}
|\langle{K^+|A_{\pi^+}|K_0^{*0}}\rangle| \simeq 0.29, 
                                          \label{eq:AME-K_0}
\end{equation}
where we have used $SU_I(2)$ isospin symmetry which is always 
assumed in this note. 

We now use this information to estimate the  decay widths  of the conventional $D_0^*$ scalar mesons. 
We tentatively assume  $m_{D_0^*}\simeq 2.35$ GeV as the mass,
which is close to the average of the experimental results 
in Refs.~\cite{BELLE-D_0,FOCUS} and is also 
compatible with the theoretical expectations mentioned before. 
The asymptotic $SU_f(4)$ symmetry relates asymptotic matrix elements 
of axial charges to each other~\cite{Hallock}, through 
relations such as
\begin{eqnarray}
&&\langle{D^+|A_{\pi^+}|D_0^{*0}}\rangle 
=2\langle{D^0|A_{\pi^0}|D_0^{*0}}\rangle    \nonumber\\
&&=\langle{D^0|A_{K^-}|D_{s0}^{*+}}\rangle 
=\langle{D^+|A_{\bar K^0}|D_{s0}^{*+}}\rangle \nonumber\\
&&=\langle{K^+|A_{\pi^+}|K_0^{*0}}\rangle.  
                                     \label{eq:AME-Api-sym}
\end{eqnarray}
Using this equation, we compare the $D_0^*\rightarrow D\pi$ 
decays with the $K_0^{*}\rightarrow K\pi$. 
Insertion of Eq.~(\ref{eq:AME-Api-sym}) and the assumed mass value for $D_0^*$ 
into Eq.~(\ref{eq:rate}) with 
Eq.~(\ref{eq:amp}) gives  
$\Gamma(D_0^{*+}\rightarrow D^0\pi^+) 
\simeq 2 \Gamma(D_0^{*+}\rightarrow D^+\pi^0) 
\simeq \Gamma(D_0^{*0}\rightarrow D^+\pi^-) 
\simeq 2 \Gamma(D_0^{*0}\rightarrow D^0\pi^0) 
\simeq 60\,\,{\rm MeV}$, 
where $SU_I(2)$ symmetry has been assumed. To allow for $SU_f(4)$ 
symmetry breaking one could reduce the above rates  by $\simeq 40\,\,\%$, but even without this correction,  
these results suggest that 
the conventional scalar $D_0^*$ mesons should be much narrower than 
the observed enhancements. Therefore, we do not accept that 
each of the measured broad bumps with a width  of $\sim 240 - 280$ MeV is 
saturated by a single state, the $D_0^*$ with a width of at most 
$\sim 100$ MeV.   This is a compelling argument for a structure consisting of at 
least two resonances, one of which is the conventional scalar 
$D_0^*\sim \{c\bar n\}$ with the width $\Gamma_{D_0^*}\sim 50-100$ 
MeV located in  the upper half of the enhancement.  For the other we propose the 
scalar four-quark meson, $\hat D_0\sim [cn][\bar u\bar d]$ with a 
narrow width, located in the lower tail of the broad bump. 

Next, we study the charm-strange scalar mesons.  We have noted that
 the experimental data on the $D_s^{*+}\gamma$ and 
$D_s^+\pi^0$ decays of the $D_{s0}^+(2317)$ favors its assignment 
to the $(I,I_3)=(1,0)$ scalar four-quark meson, 
$\hat F_I^+\sim [cn][\bar s\bar n]_{I=1}$, and we identified the  
$D_{s0}^+(2317)$ with the $\hat F_I^+$.  This opens the question ``where is the conventional $D_{s0}^{*+}\sim \{c\bar s\}$ scalar  meson?''.   Theoretical 
predictions of the mass of this 
 meson are still not stable. 
The mass of $D_{s0}^{*+}$ may be expected to be considerably  
higher than $2317$ MeV, as the non-strange $D_0^{*}$ are near or above this.  We note the results in the literature:
$m_{D_{s0}^*}\simeq 2.45 - 2.48$ GeV 
in potential models~\cite{GK,potential},  
$m_{D_{s0}^*}\simeq 2.47$ GeV,   in quenched relativistic lattice QCD~\cite{quench},
$m_{D_{s0}^*} \simeq 2.44$ GeV in unquenched ($n_f = 2$)  static lattice QCD~\cite{Bali}, and 
$m_{D_{s0}^*} \simeq 2.37$ GeV in relativistic unquenched lattice QCD~\cite{UKQCD},
 In QCDSR, the 
results are still more unstable. They are strongly dependent on 
the mass value of the charmed quark used in the calculation, i.e., for example, 
$m_{D_{s0}^*}\simeq 2.48$ GeV for $m_c\simeq 1.46$ 
GeV~\cite{QCDSR-HT} and $m_{D_{s0}^*}\simeq 2300$ GeV for 
$m_c\simeq 1.13$ GeV~\cite{Narison}.   
The latter value is close to the measured mass of the
$D_{s0}^+(2317)$. Although it may be tempting to use this agreement as evidence that the $D_{s0}^+(2317)$ is a conventional scalar meson, it is unsupported by the other mass estimates, and is hard to 
reconcile with the experimental constraint, Eq.(\ref{eq:CLEO}), 
we discussed above. 
Since there is not yet a consensus for the predicted mass of the $D_{s0}^{*}$, 
we tentatively take the value $m_{D_{s0}^*}\simeq 2.45$ GeV as expected 
from  simple quark counting with $\Delta m_s\simeq 0.1$ GeV 
and $m_{D_0^*}\simeq 2.35$ GeV. Our assignment of  the charm-strange 
scalar mesons under consideration is now 
the conventional $D_{s0}^{*+}\sim \{c\bar s\}$ with a mass 
$m_{D_{s0}^{*}}\simeq 2450$ MeV and the four-quark meson 
$\hat F_I^+$ with a mass $m_{\hat F_I}\simeq 2317$ MeV. 
The width of the $D_{s0}^+(2317)$, is now constrained to 
be $< 4.6$ MeV~\cite{PDG04}. In this note it is sufficient  that 
it is narrow, compared to typical strong decay widths.  With the  assignment of  the 
$D_{s0}^+(2317)$ to the scalar four-quark state, $\hat F_I^+$, 
such a narrow width is natural because of small overlap of the 
color and spin wavefunctions  between 
the initial charmed four-quark meson and final two pseudoscalar
meson states.  If one assigns the observed $a_0(980)$ to 
the isotriplet light four-quark meson, 
$\hat\delta^s\sim [ns][\bar n\bar s]_{I=1}$,  
as suggested long ago in Ref.~\cite{Jaffe}  and recently
in Ref.~\cite{MPPR}, 
the observed $a_0(980)\rightarrow\eta\pi$ width of about 70 MeV, gives a width of 9 MeV for $\hat F_I^+\rightarrow D_s^+\pi^0$~\cite{HT} .

To estimate the width of the conventional $D_{s0}^{*+}$ 
we return to
Eqs.(\ref{eq:rate}), (\ref{eq:amp}), (\ref{eq:AME-K_0}) 
and (\ref{eq:AME-Api-sym}), and get
\begin{equation}
\Gamma(D_{s0}^{*+}\rightarrow D^0K^+)
\simeq 36\,\,{\rm MeV},   
                                       \label{eq:width-D_{s0}^*}
\end{equation}
where the $\sim 40\,\,\%$ correction due to  asymptotic $SU_f(4)$ symmetry breaking 
has not been taken into account. The decay,  
$D_{s0}^{*+}\rightarrow D^+K^0$, also has the same rate 
because of  $SU_I(2)$ symmetry, and because these two dominate the full width of the $D_{s0}^{*+}$, we estimate the  full width
to be  
$\Gamma_{D_{s0}^{*}}\sim 70$ MeV ($\sim 40$ MeV if we allow for 
 $SU_f(4)$ symmetry breaking).
 Of course, the contribution of possible isospin non-conserving 
$D_{s0}^{*+}\rightarrow D_s^+\pi^0$ decays will be 
negligibly small.  Adapting the method of Ref.~\cite{HT} to the mass 
$m_{D_{s0}^{*}}=2450$ MeV estimates the width for this process as $1$ keV.
This  is compatible with the fact 
that no scalar resonance has been observed in the region above 
the $D_{s0}^+(2317)$ resonance up to 
$\simeq 2.7$ GeV  
in the $D_s^+\pi^0$ mass distribution~\cite{BABAR}. 
It should be noted that the CLEO collaboration~\cite{CLEO-Kubota} 
have observed a peak around $2.39$ GeV in the $DK$ mass distribution 
but it has been taken away as a false peak arising from the decay, 
$D_{s1}(2536)\rightarrow D^*K\rightarrow D[\pi^0]K$,  
where the $\pi^0$ has been missed. However, we hope that it might  
involve the true resonance corresponding to the $D_{s0}^{*+}$ 
or that the resonance could be observed by experiments with higher 
luminosity and resolution. 

To summarise,  we have studied the broad enhancements in the $D\pi$ mass 
distributions which have been independently observed by the BELLE and FOCUS 
collaborations, and have pointed out that each bump 
is unlikely to be saturated by a single scalar $\{c\bar n\}$ state. 
We expect each enhancement to have a structure including at least two peaks, one  
arising from the four-quark $\hat D\sim [cn][\bar u\bar d]$ 
and the other from  the conventional $D_0^*\sim\{c\bar n\}$, 
although the experimental collaborations have claimed that 
these bumps are consistent with the conventional scalar mesons alone. 
By comparing the decays of the $D_0^*$'s with the well-known 
$K_0^*(1430)\rightarrow K\pi$, 
the widths of the $D_0^*$'s
 predicted to be broad, $\Gamma_{D_0^*}\sim 90$ MeV 
(or $\sim 50$ MeV when the asymptotic $SU_f(4)$ symmetry breaking 
has been taken into account), but they are still not broad enough to comprise the whole bump.  In comparison,  the four-quark 
$\hat D$ mesons are expected  to have widths of at most 
$\sim 5 - 10$ MeV. 

The strange counterpart, $D_{s0}^*\sim \{c\bar s\}$, of the two quark scalar,
$D_0^*$, is expected, on the basis of many different approaches,  to have a mass around 
$m_{D_{s0}^*}\sim 2.45$ GeV.  
Its width is expected to be approximately saturated by 
the $D_{s0}^{*+}\rightarrow (DK)^+$ decays and is predicted 
to be $\sim 70$ MeV (or $\sim 40$ MeV when the asymptotic 
$SU_f(4)$ symmetry breaking has been taken into account). 

We emphasise that  the values of the masses and widths of the scalar 
resonances which have been estimated in this short note should not 
be taken too literally since the value of the width of 
the $D_{s0}^+$ which has been used as the input data in 
Ref.~\cite{Terasaki-D_s} is still somewhat uncertain,  and 
since possible mixing between $D_0^*$ and $\hat D$ through 
their common decay channels which may have considerable effects 
on the masses and widths of the mixed states~\cite{ABS} has been 
neglected. Such a mixing will depend on the details of hadron 
dynamics including four-quark mesons, and it will be a fruitful subject for
future studies. 

Finally, we point out that it is desirable that the measured broad bumps in the $D\pi$ 
mass distributions be (re)analyzed by using an amplitude 
including at least two scalar resonances. 
It is also expected that the charm-strange scalar, 
$D_{s0}^{*+}\sim\,^3P_0\,\,\{c\bar s\}$,  will be observed in 
the $DK$ channels by experiments with high  resolution and 
luminosity. 

\section*{Acknowledgments}
K.~T would like to thank the members  
of high energy  physics group, the University of Melbourne for
discussions and hospitality  during his stay. This work was 
started there. BMcK is grateful for the hospitality and support of the Yukawa Institute of Theoretical Physics during his visit there, which enabled the completion of the work.

This work is supported in part by  the 
Grant-in-Aid for Science  Research, Ministry of Education,  
Science and  Culture, Japan  (No. 13135101 and No. 16540243), and in part by the Australian Research Council grant No DP0344913.



\begin{thebibliography}{99}
\setlength{\itemsep}{2pt} 

{\normalsize 

\bibitem{BELLE-D_0} K.~Abe et al., the BELLE Collaboration,
Phys. Rev. D {\bf 69}, 112002 (2004). 

\bibitem{FOCUS} E.~W.~Vaandering, the FOCUS collaboration, 
hep-ex/0406044. 

\bibitem{BABAR} 
B.~Aubert et al., the BABAR Collaboration, 
Phys. Rev. Lett. {\bf 90}, 242001 (2003). 


\bibitem{CLEO} 
D.~Besson, the CLEO Collaboration, Phys. Rev. D {\bf 68}, 032002 
(2003). 

\bibitem{BELLE-D_s} 
P.~Krokovny et al., the BELLE Collaboration, 
Phys. Rev. Lett. {\bf 91}, 262002 (2003). 

\bibitem{DGG} A.~De~R\'ujura, H.~Georgi and S.~L.~Glashow, 
Phys. Rev. Lett. {\bf 37}, 785 (1976). 

\bibitem{chiral} 
M.~A.~Nowak, M.~Rho and
I.~Zahed, Phys. Rev. D  {\bf 48}, 4370 (1993); W.~A.~Bardeen and 
C.~T.~Hill, Phys. Rev. D {\bf 49}, 409 (1994). 


\bibitem{CH} 
H.-Y.~Cheng and W.-S.~Hou, Phys. Lett. {\bf B566}, 193 (2003). 

\bibitem{BCL} 
T.~Barnes, F.~E.~ Close and H.~J.~Lipkin,  Phys. Rev. D {\bf 68}, 
054006 (2003).  

\bibitem{atom} A.~P.~Szczepaniak, Phys. Lett. {\bf B567}, 23 (2003). 

\bibitem{BR} 
E.~van~Beveren and G.~Rupp, Phys. Rev. Lett. {\bf 91}, 012003 (2003). 

\bibitem{Lutz} E.~E.~Kolomeitzev and M.~F.~M.~Lutz, 
Phys. Lett. {\bf B582}, 39 (2003); 
J.~Hofmann and M.~F.~M.~Lutz, Nucl. Phys. A {\bf 733}, 142 (2003). 


\bibitem{MPPR} 
L.~Maiani, F.~Piccinini, A.~D.~Polosa and V.~Riquer, 
Phys. Rev. Lett. {\bf 93}, 212002 (2004); hep-ph/0412098. 

\bibitem{BPP} 
T.~E.~Browder, S.~Pakvasa and A.~A.~Petrov, Phys. Lett. {\bf 578}, 
365 (2004). 

\bibitem{Terasaki-D_s} K.~Terasaki, Phys. Rev. D {\bf 68}, 
011501(R) (2003). 


\bibitem{HT} A.~Hayashigaki and K.~Terasaki, hep-ph/0410393. 


\bibitem{GK} S.~Godfrey and R. Kokoski, Phys. Rev. D {\bf 43}, 
1679 (1991). 

\bibitem{potential} W.~Lucha and F.~F.~Schr\"oberl, Mod. Phys. Lett. 
A {\bf 18}, 2837 (2003) and references quoted therein. In 
this paper, only the $^3P_0\,\,\{c\bar s\}$ state has been studied.  

\bibitem{quench} P.~Boyle, UKQCD, Nucl. Phys. Proc. Suppl. 
{\bf 63}, 314 (1998). 

\bibitem{Bali} G.~S.~Bali, Phys. Rev. D {\bf 68}, 071501(R) (2003). 

\bibitem{UKQCD} A.~Dougall et al., the UKQCD Collaboration, 
Phys. Lett. {\bf B569}, 41 (2003). 

\bibitem{QCDSR-HT} A.~Hayashigaki and K.~Terasaki, hep-ph/0411285  
and references therein. 

\bibitem{Narison} S.~Narison, Phys. Lett. {\bf B} (in press), 
hep-ph/0307248; hep-ph/0411145 and refences therein.  

\bibitem{Terasaki-ws} K.~Terasaki, 
Soryushiron Kenkyu (Kyoto) {\bf 108}, F11 (the Proceedings of 
the YITP workshop on Progress in Particle Physics, July 22 -- 25, 
2003, Yukawa Institute for Theoretical Physics, 
Kyoto University, Kyoto), hep-ph/0309119; 
hep-ph/0309279 (in the Proceedings of the 10-th International 
Conference on Hadron Spectroscopy, Aug. 31 -- Sept. 6, 2003,  
Aschaffenberg, Germany, edited by E.~Klempt,H.~Koch and H.~Orth 
(AIP, New York, 2004), p.556. 

\bibitem{Terasaki-D_0} K.~Terasaki, hep-ph/0311069. 

\bibitem{Terasaki-mquark} K.~Terasaki, 
hep-ph/0405146 (to appear in the proceedings of the YITP workshop, 
{\it Multiquark hadrons; four, five and more?}, 
Feb. 17 -- 19, 2004, Yukawa Institute for Theoretical Physics, 
Kyoto University, Kyoto). 

\bibitem{CT} F.~E.~Close and N.~A.~T\"ornquvist, J. Phys. G {\bf 28}, 
R249 (2002). 

\bibitem{suppl} S.~Oneda and K.~Terasaki, Prog. Theor. Phys. Suppl. 
No. {82}, 1 (1985) and references quoted therein. 

\bibitem{MP} For example, V.~S.~Mathur and L.~K.~Pandit, {\it Advances 
in Particle Physics}, eds. R.~L.~Cool and R.~E.~Marshak (Interscience, 
1968), Vol. 2, p. 383 and references quoted therein. 
\bibitem{LR} H.~Leutwyler and M. Roos, Z. Phys. C {\bf 25}, 91
(1984). 

\bibitem{PDG96} Particle Data Group, R.~M.~Barnet et al., Phys. Rev. 
D {\bf 54}, 1 (1996). 

\bibitem{E687} M.~S.~Nehring, the Fermilab E687 Collaboration,
Nucl. Phys. B (Proc. Suppl.) {\bf 55A}, 131 (1997). 

\bibitem{CLEO97} J.~Bartelt et al., CLEO collaboration,
Phys. Lett. {\bf B405}, 373 (1997). 

\bibitem{HOS} H.~Hallock, S.~Oneda and M.~D.~Slaughter, Phys. Rev. 
D {\bf 15}, 884 (1977). The results are now improved by using 
the new data on the charm meson masses and the decay rate for the 
$\rho\rightarrow \pi\pi$ as the input data. 

\bibitem{CLEO-D^*} A.~Anastassov et al, the CLEO collaboration, 
Phys. Rev. D {\bf 65}, 032003 (2002). 

\bibitem{PDG04} S.~Eidelman et al., Particle Data Group, Phys. Lett. 
B {\bf 592}, 1 (2004). 

\bibitem{Hallock} H.~L.~Hallock, 
{\it $SU(3)$ in the World of More Than Three Quarks}, 
PhD theses (University of Maryland, 1978). 

\bibitem{Jaffe} R.~L.~Jaffe, Phys Rev. D {\bf 15}, 267 and 281
(1977). 

\bibitem{CLEO-Kubota} Y.~Kubota et al., the CLEO Collaboration, 
Phys. Rev. Lett. {\bf 72}, 1972 (1994). 

\bibitem{ABS} V.~V.~Anisovich, D.~V.~Bugg and A.~V.~Sarantsev,
Phys. Rev. D {\bf 58}, 111503(R) (1998). 
}

\end{thebibliography}
\end{document}